\documentclass{jpsj2}

\def\mrm{\mathrm}
\def\mbf{\mathbf}

\def\el{\mrm{el}}

\def\rr{\mbf{r}}

\def\mrm{\mathrm}
\def\mbf{\mathbf}

\def\el{\mrm{el}}

\def\rr{\mbf{r}}
\def\XX{\mbf{X}}
\def\VV{\mbf{V}}

\def\round{\partial}

\def\vphi{\varphi}

\def\el{\mrm{el}}

\def\rd2{r^{\prime2}}

\renewcommand{\d}{{\mbox{d}}}
\newcommand{\odif}[2]{{\frac{\d #1}{\d #2}}}
\newcommand{\pdif}[2]{{\frac{\partial #1}{\partial #2}}}

\newcommand{\Ds}{\displaystyle}

\title{
Abnormal diffusion of a single vortex in the two dimensional XY model
}

\author{
Tomoaki \textsc{Nogawa}\thanks{E-mail address: 
nogawa@serow.t.u-tokyo.ac.jp} 
and 
Koji \textsc{Nemoto}$^{1}$\thanks{E-mail address:
nemoto@statphys.sci.hokudai.ac.jp}
}
\inst{
Department of Applied Physics, School of Engineering,
The University of Tokyo, Bunkyo-ku, Hongo, Tokyo 113-8656
\\
$^{1}$
Department of Physics, Hokkaido University,
Sapporo, Hokkaido 060-0810 Japan
}

\abst{
We study thermal diffusion dynamics of a single vortex 
in two dimensional XY model. 
By numerical simulations we find an abnormal diffusion 
such that the mobility decreases with time $t$ as $1/\ln t$. 
In addition we construct a one dimensional diffusion-like equation 
to model the dynamics and confirm 
that it conserves quantitative property of the abnormal diffusion. 
By analyzing the reduced model, 
we find that the radius of the collectively moving region 
with the vortex core grows as $R(t) \propto t^{1/2}$. 
This suggests that the mobility of the vortex 
is described by dynamical correlation length as $1/\ln R(t)$. 
}

\kword{vortex, XY model, abnormal diffusion, Josephson junction array}

\begin{document}

\maketitle

\section{Introduction}

Vortices, which are topological defects of U(1) symmetry fields, 
plays an important role in low dimensional systems 
such as thin film super fluids, liquid crystals, 
layer superconductors and Josephson junction arrays(JJAs). 
A well known example is that two dimensional XY (2dXY) model 
exhibits a vortex driven KT phase transition \cite{Kosterlitz73} 
while the elastic theory, which ignores the vortices, 
predicts a unique phase with quasi-long-range order. 
In this case, i.e., a vortices behave as two dimensional Coulomb gas, 
which make dipole pairs in the ordered state. 

Vortex is also an important keyword 
in dynamical property of the system. 
In the out-of-equilibrium dynamics, 
relaxation process from certain initial state to the equilibrium 
at fixed temperature environment, 
it is pointed out that 
the critical relaxation of the 2dXY model seems not to be universal;  
the dynamical exponent $z$, which connects 
dynamical correlation length $L(t)$ 
and time $t$ as $L(t) \propto t^{-1/z}$, 
depends on the initial state. 
When the initial state is an ordered ground state, 
$z$ equals 2, which agrees with the result 
of the elastic description, i.e., the Gaussian model. 
On the other hand $z \approx 2.35$ \cite{Bray-Briant-Jervis00} for process 
quenching from highly disordered initial state at high temperature. 
The latter value of $z$ is, however, considered to be 
a consequence of short time correction caused by topological defects. 
In the disordered initial state the system is filled with vortices 
and vortex-antivortex pair annihilation 
is a main process of initial relaxation. 
This yields logarithmic correction as 
$L(t) \sim (t/\ln t)^{-1/2}$ 
\cite{Yurke93, Rojas1999, Bray-Briant-Jervis00}, 
which resembles $t^{-1/2.35}$ 
when the observing time is not large enough. 
The long time asymptotic behavior 
is expressed by $z=2$ as well as 
in the case of the ordered initial state.

In the phenomena mentioned above, 
interaction among vortices is important. 
On the other hand, it is also reported that 
even a single vortex causes abnormal behavior 
in transport property of a JJA system. 
Under very low magnetic field, which yields very diluted vortices, 
the frequency dependence of the vortex mobility 
behaves as $1/\omega$ 
\cite{Theron93}. 
This result conflicts with the free Coulomb gas picture 
\cite{Ambegaokar80, Shenoy85}. 
Korshunov explained this experimental observation 
(deriving corresponding facts that 
the mobility of the vortex decreases as $1/\ln t$) 
by analyzing the 2dXY model assuming an effective action 
which involves a memory kernel in the interaction term 
\cite{Korshunov94}. 
In this article, 
we study the diffusion dynamics of a single vortex based on two models 
and discuss about the origin of the memory effect. 
At first we show numerical study of the bare 2dXY model. 
Then we derive a one dimensional model 
as an approximation of the 2dXY model, 
which enables us to understand the phenomena more clearly. 
Analysis of both models reveals  
that the dynamics of a vortex shows abnormal diffusion, 
where mean square displacement grows as $t/\ln t$.

\section{Numerical analysis on the two dimensional XY model}

We study the XY spin model on a square lattice, 
whose energy is written as 
\begin{equation}
E 
= J \sum_{\langle i, j \rangle} 
\left[  1 - \cos( \theta_i - \theta_j) \right]. 
\label{eq:hamiltonian}
\end{equation}
Here $\theta_i$ indicates the angle of the XY spin at the $i$-the site 
and the summation is taken over all nearest neighbor pairs. 
The dynamics of this model is investigated 
by the overdamped Langevin equation, 
\begin{equation}
\eta \frac{d}{dt} \theta_i(t) = - J \sum_{j \in \mrm{n.n.}} 
\sin ( \theta_i(t) - \theta_j(t) ) + \zeta_i(t), 
\label{eq:eom_xy}
\end{equation}
where $\zeta_i(t)$ is a Gaussian white noise satisfying 
\begin{equation}
\langle \zeta_i \rangle = 0
\quad \mrm{and} \quad 
\langle \zeta_i(t) \zeta_j(t') \rangle = 2 \eta k_B T \delta_{ij} \delta(t-t').
\end{equation}
Here $\langle \cdots \rangle$ means the average over 
independent noise realizations. 
In the following, we set the coupling constant $J$, 
friction coefficient $\eta$ and the Boltzmann constant $k_B$ 
to unity.

Next let us introduce the quantities to observe. 
The mean square displacement (MSD) is calculated as 
\begin{equation}
D(t,t_0) \equiv \langle \left| \XX(t) - \XX(t_0) \right|^2 \rangle
\end{equation}
where $\XX(t)=(X(t),Y(t))$ is a position of a vortex at time $t$ 
and $t_0$ is a waiting time. 
The velocity auto correlation function 
is derived from $D(t,t_0)$ from the relation 
\begin{equation}
\frac{\round}{\round t} \frac{\round}{\round t_0} D(t,t_0) 
= - 2 \langle \dot{\XX}(t) \dot{\XX}(t_0)  \rangle
\equiv -2 C(t,t_0). 
\end{equation}
When the system has time translational symmetry in stationary state, 
the MSD is a function of only $t-t_0$ 
and the velocity correlation function is rewritten as 
\begin{equation}
C(t-t_0) = \frac{1}{2} \frac{\round^2}{\round t^2} D(t-t_0). 
\end{equation}

\subsection{Simulation settings}

We numerically integrated eq.~(\ref{eq:eom_xy}) 
by the second-order Runge-Kutta method \cite{Honeycutt92}. 
The samples are square-shaped including $L^2$ spins 
and open boundary condition is imposed. 
In the initial state, 
the phase $\theta_i(t=0)$ is given as 
an angle between the $x$-axis and the vector $\rr_i - \XX(0)$,  
where $\rr_i$ is a position vector of the $i$-th site 
and $\XX(0)$ is the initial position of the vortex core 
set to the center of the sample $( (L-1)/2, (L-1)/2 )$.

Next let us explain how to detect the position of the vortex core.
We calculate vorticity $n_i$ from the snap shot at time $t$ 
by summing the phase difference along the closed path 
of each plaquette as 
\begin{eqnarray}
2 \pi n_i = [ \theta_{i+\hat{x}} - \theta_i ]
  + [ \theta_{i+\hat{x}+\hat{y}} - \theta_{i+\hat{x}} ] \\
  + [ \theta_{i+\hat{y}} - \theta_{i+\hat{x}+\hat{y}} ]
  + [ \theta_i - \theta_{i+\hat{y}} ],
\end{eqnarray}
where $[x]= x - 2\pi \mrm{int }( \frac{x+\pi}{2\pi})$
and then $[x]$ returns a value between $-\pi$ and $\pi$. 
Thus $n_i$ can take 0 or $\pm 1$. 
A vortex core exists in the plaquette labeled 
by index $i$, where $n_i=1$.
The velocity of the vortex is calculated as 
$\VV(t) \equiv ( \XX(t+\tau) - \XX(t) )/ \tau$, 
where $\tau=m\Delta t$ is time interval between sequential observations 
and $\Delta t$ is an incremental time step of the simulation. 
We set to $\Delta t = 0.056$ and $m=64$. 
When $\tau$ is set to sufficiently small value, 
$\VV(t)$ equals zero in the most time steps. 
This is because the motion of the vortex is discrete and intermittent. 
In such situation 
we have to note that this velocity averaged for finite time span 
barely depends on the choice of $\tau$.

There are two difficulties in this simulation. 
One is that pairs of vortex and anti-vortex 
can be created by thermal fluctuation, 
which makes it very difficult to find the trajectory 
of the vortex initially prepared. 
Such a thermal excitation, however, is observed 
with extremely small probability at low temperature. 
We set $T = 0.250 \approx 0.28 T_\mrm{KT}$ 
where $T_\mrm{KT}$ is the critical temperature 
of the Kosterlitz-Thouless transition. 
The other problem is that 
the vortex feels attractive force from the sample edge 
so that it gets out of the sample in a certain long time. 
In order to observe the long time steady behavior 
we make an operation to keep the vortex 
around the center of the sample as follows. 
Let us consider an example case that 
the vortex core moves along the $x$(right)-direction 
by $a(\ge2)$ steps. 
At first we delete $a$ columns of spins from left edge 
and then move the all remaining spins to the left by $a$ lattice units. 
Finally we add spins to every empty columns on the right edge 
by copying the $(L-a)$-th column in the same way.
We do similar operation 
when the vortex core moves to the left, up and down. 
Such operation affects the motion of the vortex core 
to some extent but this effect will decrease 
with the system size $L$ in a systematic way.

\subsection{Numerical Result}

\begin{figure}
\begin{center}
\includegraphics[trim=10 240 0 -230,scale=0.30,clip]{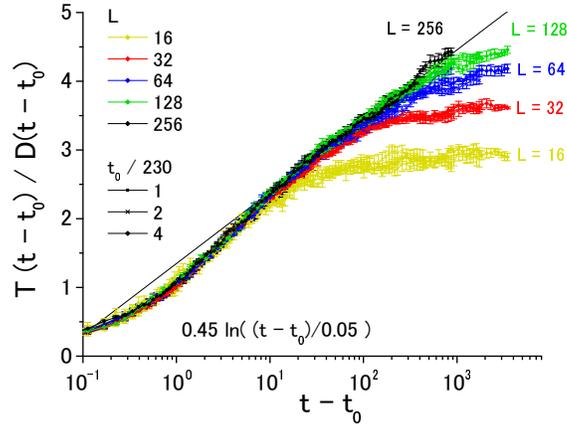}
\end{center}
\vspace{-3mm}
\caption{\label{fig:dt_xy}
Mean square displacement is plotted with $t-t_0$. 
The result for three waiting times,  
five system sizes (L=16, 32, 64, 128, 256) 
are shown together. 
Average is taken over 8192 samples. 
}
\end{figure}

If the dynamics of the vortex is the so-called ``normal diffusion'', 
the MSD would be proportional to $T t$ 
and its coefficient means the mobility. 
In fact $D(t)$ grows slower than $t$-linear behavior 
as shown in Fig.\ref{fig:dt_xy}.
In order to show the deviation from normal diffusion apparently 
we plot $T t / D(t)$, which can be regarded 
as a time-dependent effective friction coefficient 
(or inverse of the effective mobility) of the vortex.
The results for different waiting times $t_0$ are plotted together. 
Since the deviation among them is very small as far as $t-t_0 \ll t_0$, 
the system is considered to be in a stationary state. 
The {\it coefficient} is found to be proportional to $\ln t$ 
in long time regime, i.e., 
\begin{equation}
D(t) \propto T t / \ln t.
\end{equation}
Finite size effect is observed in  the long time limit.  

Figure~\ref{fig:dt_xy} shows that 
the coefficients saturate to a finite values 
which are roughly proportional 
to the logarithm of the system size $L$.

\begin{figure}
\begin{center}
\includegraphics[trim=10 240 0 -230,scale=0.30,clip]{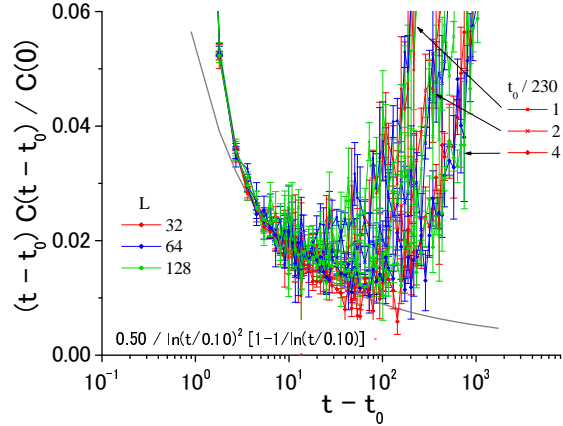}
\end{center}
\vspace{-3mm}
\caption{\label{fig:ct_xy}
Vortex velocity auto-correlation function 
multiplied by $t-t_0$. 
The divergent behavior for large $t$ 
is due to the finite observation time, 
which is set to be same length with $t_0$. 
}
\end{figure}

Next we calculate the velocity auto-correlation function 
by using the Fourier series $\VV_k$ of the $\VV(t)$ 
\begin{eqnarray}
C(n \tau) = \sum_{k=1}^{n^*-1}  |\VV_k|^2 e^{-2\pi i n k/N } \\
\VV_k = \frac{1}{\sqrt{n^*}} \sum_{n=0}^{n^*-1} 
\VV(t_0 + n \tau ) e^{2\pi i n k/N },
\end{eqnarray}
where $n^*$ is the total number of observations of the velocity 
with constant interval $\tau$. 
Although the above $C(t)$ barely depends on $\tau$, 
a normalized function $C(t)/C(0)$ does not for $t \gg \tau$. 
Note that finite time observation results 
non-vanishing constant in $C(t)$ for large $t$
(we set the observation time equivalent to the waiting time $t_0$). 
Ignoring this finite time effect 
the observation of Fig.~\ref{fig:ct_xy} supports that 
\begin{equation}
C(t) \propto \frac{d^2}{dt^2} \frac{t}{\ln t} 
= - \frac{1}{t \ln t ^2 } \left( 1 - \frac{2}{\ln t} \right). 
\end{equation}
The correlation function is negative for $t>0$. 
This is natural because 
the local motion of the vortex core driven by random force 
usually raise the interaction energy with surrounding region. 
Thus restoring force works on the vortex core. 
(Off course the system has an energy invariance 
against the global translation of spins 
if boundary effect can be ignored. )
The response time should correlate with the range of dragged region. 
For a single vortex, there is no characteristic length scale 
except the lattice unit 
and therefore the system has infinitely long-time memory.

\section{Reduced model}

In the previous section, 
we saw that a single vortex does not behave 
as a Brownian particle with normal diffusion 
but the mobility decreases as $1/\ln t$. 
For the aim to study the origin of the abnormal diffusion 
the 2dXY model is hard to analyze 
and numerical simulation is rather heavy 
(note that the observation time of the present simulations  
is not long enough to eliminate the possibility 
that $D(t) \propto t^{0.8}$, 
which is difficult to distinguish from $t/\ln t$ for small $t$). 

\begin{figure}
\begin{center}
\includegraphics[trim=150 420 160 -60,scale=0.50,clip]{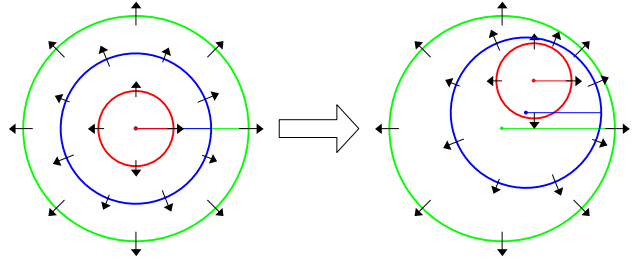}
\end{center}
\vspace{-3mm}
\caption{\label{fig:onion}
Schematic diagram of the restricted deformation 
in the reduced model. 
We consider virtual circles with radius $r$ and center $\XX(r)$. 
In fact there are infinite number of circles, 
whose radius varies continuously. 
The spins direct to the radial direction 
of the circle on which they are put on. 
The deformation is described only with the positions of the circles. 
}
\end{figure}

For this reason we propose a reduced model of this system. 
The fundamental idea is as follows. 
Considering circles with various radii 
centered on the vortex core, 
all XY spins in the energy minimal state with single vortex 
are directed the radial direction 
(see the left figure in Fig.~\ref{fig:onion}). 
We assume that the excited state 
can be described only the motion of these circles 
and spins on each circle are always along the radial direction 
(see the right figure in Fig.~\ref{fig:onion} 
and note that spins do not change there position). 
By taking the center positions as a degrees of freedom, 
the resultant equation of motion in continuum limit is 
\begin{eqnarray}
\frac{d}{dt}X(r,t) 
= 2 \left( \frac{\round^2}{\round r^2} 
- \frac{1}{r} \frac{\round}{\round r} \right) X(r,t) 
+ Z(r,t).
\nonumber \\
\langle Z(r,t) Z(r',t') \rangle
= 2 \frac{r}{\pi} T \delta(r-r') \delta(t-t'). 
\label{eq:eom_1d}
\end{eqnarray}
The detail of the derivation is 
shown in the appendix. 
Here $X(r,t)$ is the $x$-component of the 
center of the circle with radius $r$. 
The $y$-component obeys the same equation 
and decoupled with $X(r,t)$. 
The position of the vortex core is 
identified with $\lim_{r \rightarrow 0}\XX(r,t)$.

\section{Numerical integration of reduced model}

To confirm the validity of the above one dimensional model 
we numerically integrate the equation of motion. 
We write the elastic energy 
\begin{equation}
E_\mrm{el} = \frac{1}{2} \sum_j \frac{ |X_{j+1} - X_j|^2 }{r_j}
= \frac{1}{2} \sum_j \frac{ \Delta X_j^2 }{r_j}
\end{equation}
as a discrete version of eq.~(\ref{eq:eom_1d}) 
where $X_i(t)\quad(i=0,1,2,\cdots,L-1)$
is a degree of freedom on the lattice points 
and $\Delta X_i = X_{i+1} - X_i$. 
We use a system with reflective symmetry, 
i.e., 
$r_j=j+1/2$ for $j \le L-1$ 
and $r_j=L-j-1/2$ for $j \ge L$.
Therefore both of $X_0$ and $X_{L-1}$ represents the 
position of the vortex core. 
The Langevin equation is written as 
\begin{eqnarray}
\frac{d}{dt} X_i
&=& - i \frac{\round E_\mrm{el}}{\round X_i} + Z_i (t)
\nonumber \\
&=& \frac{1}{1-1/4 i^2} 
\left[
\left( \Delta X_{i} - \Delta X_{i-1} \right) 
- \frac{ \Delta X_{i} + \Delta X_{i-1} }{2 i} 
\right] 
+ Z_i (t). 
\label{eq:eom_1dnum}
\end{eqnarray}
where 
\begin{equation}
\langle Z_i(t) Z_j(t') \rangle = 2 T i \delta_{ij} \delta(t-t')
\end{equation}
In this equation temperature can be absorbed 
by scaling $X$ with $\sqrt{T}$ and therefore we set $T=1$.
We set $\Delta X_{-1} = \Delta X_{L-1} = 0$ 
as an open boundary condition. 
Actually we approximate the denominator in eq.~(\ref{eq:eom_1dnum}), 
$1 - 1/4i^2 $, with unity. 

\subsection{Abnormal diffusion}

\begin{figure}
\begin{center}
\includegraphics[trim=10 240 0 -230,scale=0.30,clip]{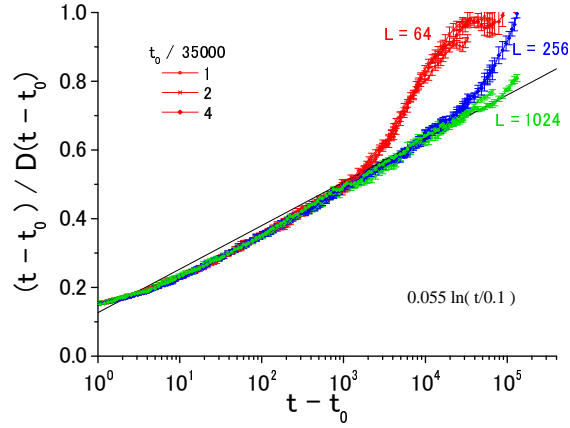}
\end{center}
\vspace{-3mm}
\caption{\label{fig:dt_1d}
Mean square displacement of the vortex center $X_0(t)$ 
for the reduced one dimensional model.
}
\end{figure}

\begin{figure}
\begin{center}
\includegraphics[trim=10 240 0 -230,scale=0.30,clip]{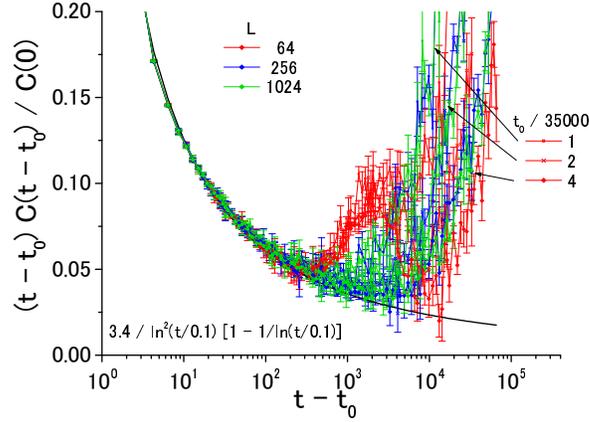}
\end{center}
\vspace{-3mm}
\caption{\label{fig:ct_1d}
Vortex velocity auto-correlation function 
multiplied by $t-t_0$ 
for the reduced one dimensional model. 
}
\end{figure}

In order to compare the present reduced model 
with the original two dimensional XY model, 
we calculate the MSD and velocity auto correlation function 
of $X_0(t)$ and $X_{L-1}(t)$. 
Figure \ref{fig:dt_1d} and \ref{fig:ct_1d} is a result 
of numerical calculation. 
The behaviors agree with those of the two dimensional XY model; 
logarithmic correction to the normal diffusion is observed. 
It can be said that the present model holds the essential property 
of the abnormal diffusion of the original model. 
Furthermore we can perform much longer time simulation 
on this model than on the 2dXY model 
and observe logarithmic property more clearly.

\subsection{Dynamical correlation length}

\begin{figure}
\begin{center}
\includegraphics[trim=10 240 0 -230,scale=0.30,clip]{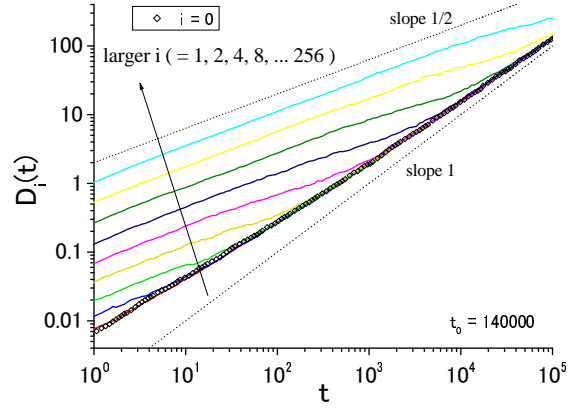}
\end{center}
\vspace{-3mm}
\caption{\label{fig:dit_1d}
Mean square displacement of 
sites $i = 0, 2^0, 2^1, \cdot, 2^9$ are shown together.
}
\end{figure}

On the reduced model 
we can easily investigate the behavior 
of off-core region. 
The MSD of $i$-th site 
\begin{equation}
D_i(t-t_0) = \langle | X_i(t) - X_i(t_0) |^2 \rangle
\end{equation}
is shown in Fig.~\ref{fig:dit_1d}. 
For small $t$, 
$D_i(t)$ is proportional to $t^{1/2}$. 
This is a standard behavior of an stochastic diffusion equation,
\begin{equation}
\frac{\round }{\round t} X(x,t) = 
\frac{\round^2 }{\round x^2} X(x,t)+ Z(x,t), 
\end{equation}
which lacks the gradient term in eq.~(\ref{eq:eom_1d}). 
Since local temperature increases with $i$, 
the initial coefficient of $t^{1/2}$ term does as well. 
The core region, however, shows different behavior. 
Seeing in logarithmic scale the growth rate is larger 
than that for off-core region. 
This ease to move is due to the weak confinement 
in the vicinity of the free edge on the one side. 
After $D_0(t)=\langle X_0(t)^2 \rangle$ 
catches up with $D_i(t)$, $D_i(t)$ coincides with $D_0(t)$.  
As time goes by, 
more and more regions moves with the core. 
Therewith the growth $D_0(t)$ becomes slower with factor $1/\ln t$ 
(seems to become faster in double-logarithmic scale). 
This suggests that the dynamics of vortex is a collective one 
and the mobility becomes smaller as its effective radius 
of the collective motion becomes large. 

\begin{figure}
\begin{center}
\includegraphics[trim=10 240 0 -230,scale=0.30,clip]{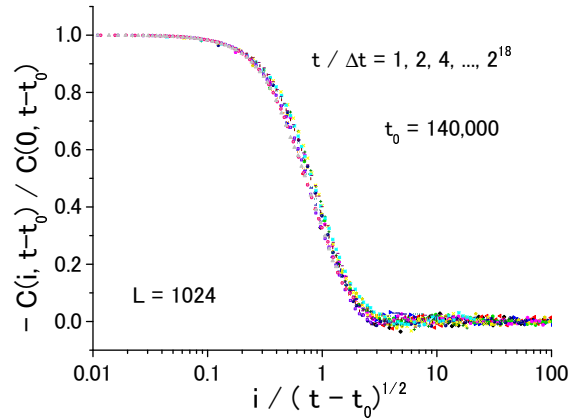}
\end{center}
\vspace{-3mm}
\caption{\label{fig:cit_1d}
The strain correlation function is plotted as a function of 
site $i$. 
The data at different time is plotted together, 
which collapse to a universal curve 
by scaling horizontal axis by $\sqrt{t}$.
}
\end{figure}

The range of collective motion can be estimated by 
the correlation function; 
\begin{equation}
C(i,t) = \frac{\langle X_0(t) X_i(t) \rangle}{ \sqrt{D_0(t) D_i(t)} }.
\end{equation}
A universal scaling function is found 
so that 
\begin{equation}
C(i,t) = F\left( i / R(t) \right) \approx e^{- i / R(t)}
\end{equation}
where 
\begin{equation}
R(t) \propto t^{1/2}. 
\end{equation}
In Fig.~\ref{fig:cit_1d} $C(i,t)$ is plotted as a function 
of $i$ scaled by $t^{1/2}$.

\section{Discussions}

We have shown that 
a single vortex exhibits abnormal diffusion 
even though there is no inter-vortex interaction.
What is important is that vortex is not a mere point particle 
but its motion drags the phase field 
of the surrounding region. 
As a natural result velocity auto-correlation function 
has memory with negative correlation. 
For an isolated vortex, 
the influence of its core motion spreads infinitely large range 
and then correlation time also diverges. 

Based on the reduced one-dimensional model, 
clearer picture of the phenomena is obtained. 
In addition, we can perform numerical simulations 
for large scale both in time and space. 
We found that the correlation length $R(t)$ 
grows as $t^{1/2}$ not as $(t/\ln t)^{1/2}$. 
This means that the logarithmic correlation 
does not come from the growth law of the correlation length $R(t)$ itself 
but originates with the size dependence of the mobility as $1/\ln R(t)$. 
This is consistent with the finite size effect of the mobility 
in the 2dXY model. 

The reduced model, eq.~(\ref{eq:eom_1d}), has two points 
that ordinary stochastic diffusion equation does not have. 
Those are the gradient term and the position dependence of temperature. 
We found that the $i$-linear dependence of the local temperature 
is not essential for the logarithmic correction 
but the gradient term in eq.~(\ref{eq:eom_1d}) is important 
(not shown here). 
This term has a function 
to make the deformation amplitude $X(r)$ propagate 
to the positive direction of $r$. 
This makes the diffusion of the end of chain slower 
than that of the ordinary elastic chain. 
This is in contrast with the Bessel equation with $n=0$ 
which has positive sign on the gradient term. 
Since the derivative kernel of the eq.~(\ref{eq:bessel}) is 
that of the 1st Bessel equation (see the appendix), 
expansion of the solution with the Bessel functions will be useful 
and the exact analysis of the present reduced model may be possible. 
This is challenging open question.

\acknowledgments

The present work is supported by 21st Century COE program 
``Topological Science and Technology'' 
and the Ministry of Education, Science, Sports and Culture, 
Grant-in-Aid for Young Scientists (A), 19740227, 2007.
A part of the computation in this work has been done 
using the facilities of the Supercomputer Center, 
Institute for Solid State Physics, University of Tokyo.

\appendix
\label{sec:appendix}

\section{Derivation of the reduced model}

\subsection{Elastic energy of single vortex}

At first we consider the interaction energy 
of the two dimensional XY model, eq.~(\ref{eq:hamiltonian}).
In the elastic continuum approximation, 
which is justified for the region away from vortex core, 
the energy is written as 
\begin{equation}
E = \int dr^2 \frac{1}{2} | \nabla \theta(\rr) |^2, 
\label{eq:eng_int}
\end{equation}
where $\rr=(x,y)$. 
A metastable state having an isolated vortex is written as 
\begin{equation}
\theta_0(\rr) = \arctan(y/x). 
\label{eq:theta0}
\end{equation}
(Strictly speaking we have to add or subtract $\pi$ for $x<0$ 
since the arctangent function returns the value 
between $-\pi/2$ and $\pi/2$.) 
The elastic energy of this state is 
\begin{equation}
E_0 = \frac{1}{2} \int_a^R dr 2\pi r \left( \frac{1}{r} \right)^2 
= \pi \ln \frac{R}{a}, 
\label{eq:e0}
\end{equation}
where $a$ is an ultraviolet cut-off length. 

We evaluate the integral in eq.~(\ref{eq:eng_int}) supposing 
\begin{equation}
\theta(\rr)=\theta_0(\rr')
\end{equation}
and $\rr'$ is related to $\rr$ as
\begin{equation}
\rr=\rr'+\XX(r'),\quad r'=|\rr'|.
\end{equation}
To transform the integral variable from $\rr$ to $\rr'$
we first evaluate the Jacobian $J$ to scale $\d^2r=J\d^2r'$:
\begin{equation}
J=\frac{\partial(x,y)}{\partial(x',y')}=\det
\begin{pmatrix} 1 + \hat{x}'X' & \hat{y}'X' \\ \hat{x}'Y' & 1 + \hat{y}'Y'\end{pmatrix}
=1+\hat{\rr}'\cdot\XX',
\end{equation}
where $\Ds \XX'=\odif{\XX}{r'}$, and 
$\hat{\rr'}=(\hat{x}'=x'/r', \hat{y}'=y'/r')$
is the unit vector parallel to $\rr'$.
To evaluate the gradient
\begin{equation}
\nabla\theta(\rr)=
\begin{pmatrix}
\pdif{x'}{x} & \pdif{y'}{x} \\
\pdif{x'}{y} & \pdif{y'}{y}
\end{pmatrix}
\nabla'\theta_0(\rr'),
\end{equation}
we need the transform matrix (the Hessian)
\begin{equation}
\begin{pmatrix}
\pdif{x'}{x} & \pdif{y'}{x} \\
\pdif{x'}{y} & \pdif{y'}{y}
\end{pmatrix}
=
\begin{pmatrix}
\pdif{x}{x'} & \pdif{y}{x'} \\
\pdif{x}{y'} & \pdif{y}{y'}
\end{pmatrix}^{-1}
=\frac{1}{J}
\begin{pmatrix} 1 + \hat{y}'Y' & -\hat{x}'Y' \\ -\hat{y}'X' & 1 + \hat{x}'X'\end{pmatrix}.
\end{equation}
Knowing the gradient with respect to $\rr'$
\begin{equation}
\nabla'\theta_0(\rr')=\frac{1}{r'}
\begin{pmatrix}
-\hat{y}' \\ \hat{x}'
\end{pmatrix}
\end{equation}
we obtain
\begin{equation}
\nabla\theta(\rr)=
\frac{1}{Jr'}
\begin{pmatrix}-\hat{y}'-Y' \\ \hat{x}' + X'\end{pmatrix},
\end{equation}
and the integrand of $E$ reads
\begin{eqnarray}
\d^2r|\nabla\theta(\rr)|^2&=&
{\d^2r'}\frac{1+2\hat{\rr}'\cdot\XX'+|\XX'|^2}{Jr^{'2}}
\nonumber \\
&=&
{\d^2r'}\frac{1+\hat{\rr}'\cdot\XX'+|\XX'|^2-(\hat{\rr}'\cdot\XX')^2+O(|\XX'|^3)}{r^{'2}}.
\end{eqnarray}
The integration of the first term corresponds to the undistorted vertex energy $E_0$
while that of the second term vanishes 
due to the rotational symmetry with respect to $\rr'$.
Thus we finally find
\begin{equation}
E_{\rm el}=E-E_0
=\frac{1}{2}\int_a^R\d r'\pi\frac{|\XX'|^2}{r^{'2}}+O(|\XX'|^3).
\label{eq:eng_int2}
\end{equation}Note that two degrees of freedom 
$X(r)$ and $Y(r)$ are decoupled. 
Therefore all we have to treat is 
one component field $X(r)$ with one dimensional parameter $r$.

By taking variation of the energy function in eq.~(\ref{eq:eng_int2}) 
we obtain the energy minimal condition  
\begin{eqnarray}
\frac{\delta E_\el}{\delta X(r)} 
= - 2\pi \round_r \frac{1}{r} \round_r X(r)
= 0.
\end{eqnarray}
On the boundary condition; $X(0)=X_0$ and $X(R)=0$, 
the solution is obtained as 
\begin{equation}
X(r) = X_0 \left[ 1 - \left( \frac{r}{R} \right)^2 \right]. 
\end{equation}

\subsection{Energy dissipation}

On the next step, we derive effective friction force for $\XX$'s. 
By using phase variable the energy dissipation rate 
of the whole system can be written as 
\begin{equation}
\frac{dE_\mrm{el}}{dt} = -\int d r^2 |\dot\theta(\rr,t)|^2. 
\end{equation}
Remembering $\tan \theta(\rr) = 
\left( y-Y\right)/\left( x-X \right)$, 
we obtain 
\begin{eqnarray}
\frac{\round \theta(\rr)}{\round X} = \frac{\sin \theta(\rr)}{r'}
,\quad
\frac{\round \theta(\rr)}{\round Y} = \frac{\cos \theta(\rr)}{r'}
\end{eqnarray}
By using this,  
\begin{eqnarray}
\frac{d E}{d t} 
&=& - \int dx dy \left| 
\frac{d X}{dt} \frac{\round \theta}{\round X} + 
\frac{d Y}{dt} \frac{\round \theta}{\round Y} 
\right|^2 
\nonumber \\
&=& - \int r' dr' \frac{1}{\rd2}  
\int_0^{2\pi} d\vphi' 
\left| \frac{dX}{dt}  \right|^2 \sin^2 \vphi' 
+ \cdots 
+ O(X^3)
\nonumber \\
&=& - \int dr' \pi \frac{1}{r'} \left( 
\left| \frac{dX}{dt} \right|^2 + 
\left| \frac{dY}{dt} \right|^2
\right) 
+ O(X^3). 
\end{eqnarray}
Again $X$ and $Y$ are decoupled. 
Therefore friction force acting on the region $(r', r'+dr')$ 
is 
\begin{equation}
- \frac{dr'}{r'} \left( 
\frac{dX}{dt}, \frac{dY}{dt} 
\right)
\end{equation}

\subsection{Equation of motion}

We can construct overdamped equation of motion for $\XX(r,t)$ 
by considering local balance between the elastic and friction forces,  
\begin{eqnarray}
\frac{\pi}{r} \frac{d}{dt}X(r,t) 
= 2 \pi \round_r \frac{1}{r} \round_r X(r,t).
\nonumber \\
\frac{d}{dt}X(r,t) 
= 2 \left( \frac{\round^2}{\round r^2} 
- \frac{1}{r} \frac{\round}{\round r} \right) X(r,t).
\end{eqnarray}

By putting $u(r,t)=X(r,t)/r$, we obtain 
\begin{eqnarray}
\frac{d}{dt}u(r,t) = 
\left( \frac{\round^2}{\round r^2} 
+ \frac{1}{r} \frac{\round}{\round r} 
- \frac{1}{r^2} 
\right) u(r,t). 
\label{eq:bessel}
\end{eqnarray}
Assuming separable solution $u(r,t) = \Upsilon(kr) \exp(-k^2 t)$, 
we obtain the Bessel equation 
\begin{eqnarray}
\left( \frac{\round^2}{\round x^2} 
+ \frac{1}{x} \frac{\round}{\round x} 
+ 1 - \frac{n^2}{x^2} 
\right) \Upsilon(x) = 0
\end{eqnarray}
with $n=1$. 
It is natural that deformation is not isotropic ($n=0$) 
but anisotropic ($n=1$) 
since the translational motion of vortex breaks 
the circular symmetry.

\subsection{Thermal noise}

Finally we consider the property of thermal noise. 
We introduce Gaussian white noise $Z(r,t)$, 
\begin{eqnarray}
\frac{d}{dt}X(r,t) = - \frac{r}{\pi} 
\frac{\delta E(\{X\})}{\delta X(\rr,t)} + Z(r,t).
\end{eqnarray}
This force $Z(r,t)$ have to satisfy the fluctuation-dissipation 
relation 
\begin{equation}
\langle Z(r,t) Z(r',t') \rangle
= 2 \frac{r}{\pi} T \delta(r-r') \delta(t-t')
\end{equation}
to realize the canonical distribution at temperature $T$ 
in equilibrium. 
The deviation of this random force is not uniform in space 
but proportional to $r$. 
It can be said that effective temperature becomes higher with $r$.


\end{document}